\documentclass[%
showpacs,
 amsmath,amssymb,
 aps,
 prl,
 10pt,
 twocolumn,
 floats,
]{revtex4-1}

\usepackage[latin1]{inputenc}
\usepackage{graphics}
\usepackage{dcolumn}
\usepackage{bm}
\usepackage{units}
\usepackage{SIunits}
\usepackage{color}

\newcommand{\ramSymb}{\mathrm{Ram}}
\newcommand{\seSymb}{\mathrm{SE}}

\newcommand{\pEc}{p_e^{(\seSymb)}}
\newcommand{\pRm}{p_e^{(\ramSymb)}}

\newcommand{\ram}[1]{{{#1}^{(\ramSymb)}}}
\newcommand{\ramsub}[1]{{{#1}_\mathrm{R}}}
\newcommand{\se}[1]{{{#1}^{(\seSymb)}}}
\newcommand{\sesub}[1]{{{#1}_\seSymb}}

\newcommand{\ttwo}{T_2^\prime}
\newcommand{\ttwos}{T_2^\ast}
\newcommand{\techo}{ t_\mathrm{echo}}

\newcommand{\nato}{\alpha}

\newcommand{\ntot}{N_\mathrm{load}}
\newcommand{\Boff}{B_\mathrm{off}}

\newcommand{\td}{\tau}

\newcommand{\ket}[1]{ \left|#1\right\rangle}
\newcommand{\bra}[1]{ \left\langle#1\right|}
\newcommand{\braket}[1]{\left|#1\right\rangle\!\left\langle#1\right|}

\newcommand{\dls}{ \bar{\delta}_\mathrm{ls} }
\newcommand{\dmw}{{\delta}_\mathrm{MW} }

\newcommand{\Gauss}{\mathrm{G}}
\newcommand{\ud}{\mathrm{d}}
\newcommand{\cheat}{C_\mathrm{h}}

\newcommand{\inirho}{\rho_0}
\newcommand{\op}{\mathcal{R}}
\newcommand{\seqOp}{\mathcal{M}}

\newcommand{\evol}{\mathcal{U}}
\newcommand{\free}[1]{\mathop{\evol_0\!\left(#1\right)}}

\newcommand{\decay}{\boldsymbol{\mathcal{C}}}

\begin{document}


\title{Coherence properties of nanofiber-trapped cesium atoms}

\author{D. Reitz}
\altaffiliation{These authors contributed equally.}
\author{C. Sayrin}
\altaffiliation{These authors contributed equally.}
\author{R. Mitsch}
\author{P. Schneeweiss}
\author{A. Rauschenbeutel}
\email{Arno.Rauschenbeutel@ati.ac.at}
\affiliation{%
 Vienna Center for Quantum Science and Technology,\\
 TU Wien -- Atominstitut, Stadionallee 2, 1020 Vienna, Austria
}%

\date{\today}

\begin{abstract}
We experimentally study the ground state coherence properties of cesium atoms in a nanofiber-based two-color dipole trap, localized $\sim 200\,\nano\meter$ away from the fiber surface. Using microwave radiation to coherently drive the clock transition, we record Ramsey fringes as well as spin echo signals and infer a reversible dephasing time of $\ttwos=0.6\,\milli\second$ and an irreversible dephasing time of $\ttwo=3.7\,\milli\second$. By modeling the signals, we find that, for our experimental parameters, $\ttwos$ and $\ttwo$ are limited by the finite initial temperature of the atomic ensemble and the heating rate, respectively. Our results represent a fundamental step towards establishing nanofiber-based traps for cold atoms as a building block in an optical fiber quantum network.

\end{abstract}
\maketitle




Over the past years, hybrid quantum systems have attracted considerable attention~\cite{Wallquist2009}. In the specific case of light--matter quantum interfaces~\cite{Lvovsky09,Simon10,Stute13}, they combine the advantages of photons for transmitting quantum information and of long-coherence-time systems, such as dopant ions in crystals, NV centers, quantum dots, single trapped neutral atoms and ions, and atomic ensembles, for storing and processing quantum information and for realizing long-distance quantum communication~\cite{Duan2001}. In the context of quantum networks~\cite{Kimble08}, it would be highly desirable to connect these matter-based storage and processing units via optical fiber links. A promising approach towards the realization of such fiber-based quantum interfaces consists in coupling cold neutral atoms to photonic crystal fibers~\cite{Christensen08,Bajcsy09,Vorrath10}. Another technique with high potential involves trapping and interfacing cold atoms in the evanescent field surrounding optical nanofibers. Using the optical dipole force exerted by a blue- and a red-detuned nanofiber-guided light field~\cite{Dowling96,LeKien04}, two-color traps have been demonstrated experimentally with laser-cooled cesium atoms~\cite{Vetsch10,Goban12}.

In order to implement quantum protocols with atoms coupled to nanophotonic devices, good coherence properties are a  prerequisite but cannot be taken for granted:  various effects, like Johnson noise~\cite{Henkel99} or patch potentials~\cite{McGuirk04} may occur and hamper long coherence times~\cite{Lacroute12}. When coupling to optical near-fields, this is all the more critical because of the small atom--surface distance of typically a few hundred nanometers. Here, using Ramsey interferometry as well as spin-echo techniques, we measure, to the best of our knowledge for the first time, the reversible and irreversible dephasing times of atoms in such an environment. Specifically, we experimentally characterize and model the ground state coherence of the clock transition of cesium atoms stored in the nanofiber-based two-color trap realized in~\cite{Vetsch10}. Remarkably, the inferred coherence times extend up to milliseconds. 

\begin{figure}
	\includegraphics{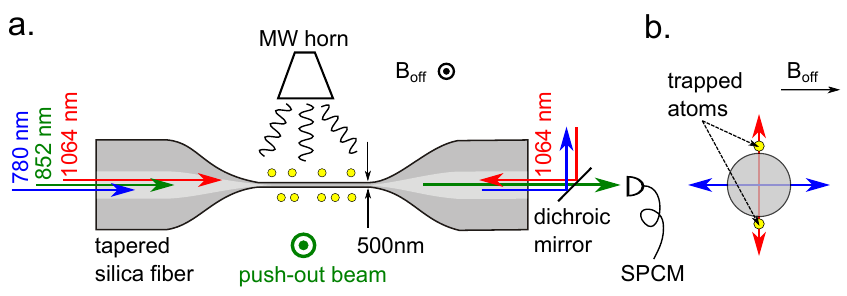}
	\caption{{\bf a.} Sketch of the experimental set-up including the tapered optical fiber, the trapping, probe and push-out laser fields, the microwave antenna, and the single-photon counter (SPCM). {\bf b.} Orientation of the plane of the quasi-linear polarizations of the blue- and red-detuned trapping fields, the atoms, and the magnetic offset field in a plane transversal to the fiber axis. }
\label{fig:setup} 
\end{figure}

The experimental set-up is sketched in Fig.~\ref{fig:setup}a and is described in detail in~\cite{Vetsch10,Vetsch12}. Cesium atoms are trapped in the evanescent field surrounding the nanofiber-waist of a tapered optical fiber. The atoms are located about $200\,\nano\meter$ above the nanofiber surface in two diametric one-dimensional arrays of potential wells, with at most one atom per trapping site. Localization in the three (radial, azimuthal and axial) directions is achieved with trap frequencies of $(200, 140, 315)\,\kilo\hertz$. In order to drive transitions between the hyperfine ground states of the trapped atoms, we use a tunable microwave (MW) field at a frequency of $9.2\,\giga\hertz$. In the following, we limit our study to the so-called clock transition between the states $\ket{e}\equiv\ket{6S_{1/2}, F=4,m_{\rm F}=0}$ and $\ket{g}\equiv\ket{6S_{1/2},F=3,m_{\rm F}=0}$. This $\ket{g} \to \ket{e}$ transition only exhibits a quadratic Zeeman shift and can be selectively addressed by applying a homogeneous magnetic offset field, $\Boff$. The latter is oriented perpendicular to the plane containing the atoms, see Fig.~\ref{fig:setup}b. 

We load the cesium atoms from a magneto-optical trap into the nanofiber trap via an optical molasses stage~\cite{Vetsch10}. In order to then prepare an ensemble of atoms in $\ket{g}$, we perform a state purification sequence: first, all atoms are optically pumped into the $F=4$ hyperfine ground state, then the atoms in $\ket{e}$ are selectively transferred to $\ket{g}$ by a MW pulse, and, finally, the remaining atoms in $F=4$ are removed from the trap by means of a $\sigma^+$-polarized ``push-out'' laser beam resonant with the light-shifted $F=4 \to F'=5$ transition of the cesium D2 line~\cite{Kuhr05}. We characterized the efficiency of this purification procedure and found that at least 62~\% of the remaining atoms end up in state $\ket{g}$ while the population in the $F=4$ manifold is negligible. In our nanofiber-trap, spin flips from $\ket{g}$ to adjacent Zeeman substates may occur because the atoms move in a light-induced fictitious magnetic field that exhibits a strong gradient of several Gauss per micrometer~\cite{Sukumar97,ToBePublished}. We observed, however, that these spin flips are suppressed for a magnetic offset field of $\Boff \geq 3\,\Gauss$. Therefore, a field of $3\,\Gauss$ is applied for all coherent manipulations of the atomic spins.

\begin{figure}
	\includegraphics{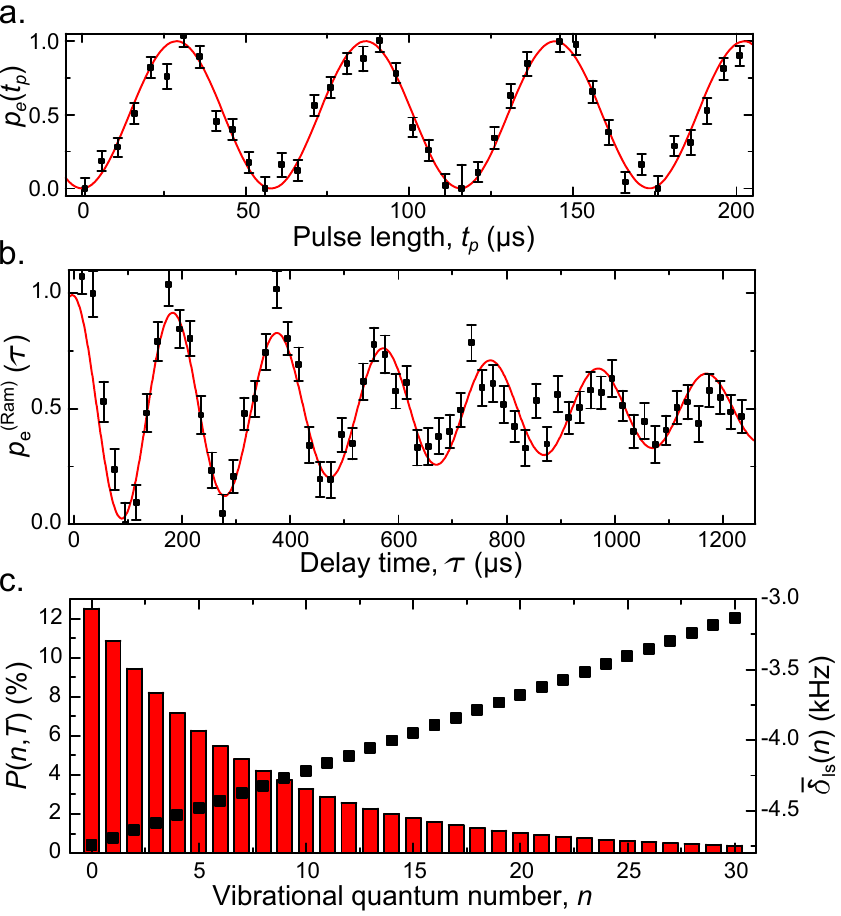}
	\caption{{\bf a.} Rabi oscillations: Probability $p_e(t_{\rm p})$ for an atom to be in state $\ket{e}$ after a resonant MW pulse of length $t_{\rm p}$. The data points correspond to an average over 80 experimental realizations (scaling factor $\alpha = 0.015$). The solid line is the result of a global fitting routine (see text). {\bf b.} Ramsey interferometry: Data averaging and fitting procedure like in {\bf a.}; the probability $\pRm(\td)$ is plotted as function of the time interval $\td$ between the $\pi/2$ pulses (scaling factor $\alpha = 0.018$). {\bf c.} Vibrational state-dependent differential light-shift $\dls(n)$ (squares) and histogram of the phonon number probability distribution $P(n,T)$ calculated at $T=71\,\micro\kelvin$.}
\label{fig:RabiRamsey}
\end{figure}

We first record Rabi oscillations between states $\ket{g}$ and $\ket{e}$ by applying a resonant MW pulse of variable duration. The transfer probability $p_e=N_e/N_{\rm tot}$ is given by the ratio of the number of atoms that were transferred to $\ket{e}$, denoted $N_e$, and of the sum of the numbers of atoms in $\ket{e}$ and $\ket{g}$, denoted $N_{\rm tot}$. Given the fact that the $F=4$ manifold is depleted after the purification and that the MW selectively drives the $\ket{g} \to \ket{e}$ transition, we identify $N_e$ with the number of atoms in $F=4$ and measure the latter by recording the transmission spectrum of a quasi-linearly polarized fiber-guided light field scanned over the light-shifted $F=4 \to F'=5$ transition of the D2 line~\cite{Vetsch10} after setting $\Boff$ to zero. Because of shot-to-shot fluctuations of $N_{\rm tot}$, the latter has to be determined for each experimental run. However, we do not have direct access to this quantity because the probing on the $F=4 \to F'=5$ transition leads both to loss of atoms from the trap and to transfer of atoms from $\ket{e}$ to $\ket{g}$ by Raman scattering. Instead, we measure the number of atoms in the trap prior to the state purification procedure, $\ntot$, which is proportional to $N_{\rm tot}$. The ratio $\nato=N_{\rm tot}/\ntot$ is found to be stable and is determined from the  analysis presented below.

The inferred probability $p_e=N_e/(\nato\ntot)$ for an atom initially in state $\ket{g}$ to be in $\ket{e}$ after the MW pulse is shown in Fig.~\ref{fig:RabiRamsey}a. The observed Rabi oscillations (resonant Rabi frequency $\Omega_0/2\pi \simeq 17$~kHz) exhibit no visible damping over $200~\micro\second$. Longer measurements (not shown) reveal an exponential reduction of the amplitude of the oscillations which decay to $p_e\simeq 0.5$ in a characteristic time of $(3.4\pm0.2)\,\milli\second$. This damping originates from fluctuations of the power of the MW source.

In order to study the coherence properties of the atomic ensemble during free evolution, Ramsey interferometry is performed: Two successive $\pi/2$-pulses, separated by a  time $\td$, are applied and the transfer probability $\pRm(\td)$ is measured, see Fig.~\ref{fig:RabiRamsey}b. The frequency of the MW source is detuned by $-5\,\kilo\hertz$ with respect to the frequency used to record the Rabi oscillations. The observed Ramsey fringes reveal a coherence time of several $100\,\micro\second$. Their decay can be explained by inhomogeneous broadening of the clock transition which is caused by the finite temperature of the atomic ensemble in combination with the dependence of the transition frequency on the vibrational state of the atoms due to differential light-shifts~\cite{Kuhr05}. These differential light-shifts are proportional to the intensity of the trapping fields and thus depend on the position $\vec{r}$ of the atom in the trap. We denote the frequency of the clock transition at position $\vec{r}$ as $\omega(\vec{r})$ and the corresponding differential light-shift as $\delta_{\rm ls}(\vec{r}) = \omega(\vec{r})-\omega_{0}$, where $\omega_{0}$ is the transition frequency in free space. The variation of $\delta_{\rm ls}(\vec{r})$ in the azimuthal and axial trapping directions is significantly smaller than in the radial direction. Therefore, in the following, only the radial dependence is taken into account. For the calculation of the motional eigenfunctions, we make the simplifying assumption that the motions in the three directions are decoupled and take the strong anharmonicity in the radial direction into account~\cite{LeKien04}. An atom in a certain radial vibrational state $\ket{n}$ will experience the differential light-shift $\dls(n) = \bra{n}\delta_{\rm ls}(\hat{r})\ket{n}$, see Fig.~\ref{fig:RabiRamsey}c, where $n$ is the phonon number and the trap parameters are the same as in~\cite{Vetsch10}. 

Thus, the density matrix of the atomic pseudo spin-1/2 just before probing, calculated in the MW field rotating frame, depends on $n$ and reads
\begin{eqnarray}
	\ram\rho(t;n) &=& \ramsub{\seqOp}(t;n)\,\inirho\,\ramsub{\seqOp}^\dag(t;n),\label{eq:defRho} \\
	\ramsub{\seqOp}(t;n) &=&\op_{\pi/2}\,\free{t;n}\,\op_{\pi/2}, \nonumber	\\
	\free{t;n} &=& \exp\left[-i(\dls(n)-\dmw)\,t\,\sigma_z/2\right], \nonumber	
\end{eqnarray}
where $\sigma_z$ is the Pauli matrix, $\inirho=\braket{g}$, and $\dmw=\omega_\mathrm{MW}-\omega_{0}$ is the detuning of the MW field with respect to the clock transition in free space. Here, $\ramsub\seqOp(t;n)$ is the evolution operator corresponding to the total experimental sequence, while $\free{t;n}$ is the free evolution operator. The operator $\op_\theta$ corresponds to a rotation of the atomic spin by the angle $\theta$ about the $y$-axis of the Bloch sphere. We assume a Boltzmann distribution for the initial occupation probability $P(n,T)$ of the vibrational states which is plotted in Fig.~\ref{fig:RabiRamsey}c for a temperature of $T=71\,\micro\kelvin$. For a given $P(n,T)$, the probability $\pRm(\td)$ of detecting an atom in $\ket{e}$ for a time delay $\td $ then reads
\begin{align}
\pRm(\td ) &= \sum_n P(n,T) \,\rho^{(\ramSymb)}_{ee}(\td ;n) \label{eq:idealProb}~.
\end{align}
In Fig.~\ref{fig:RabiRamsey}b, we plot $\pRm(\td )$ for $T=71\,\micro\kelvin$ and find that our model describes the experimental data well. We define the reversible dephasing time $\ttwos$ as the time at which the amplitude of the Ramsey fringes is reduced by $50\%$, yielding $\ttwos = 0.6\,\milli\second$.

\begin{figure}
	\includegraphics{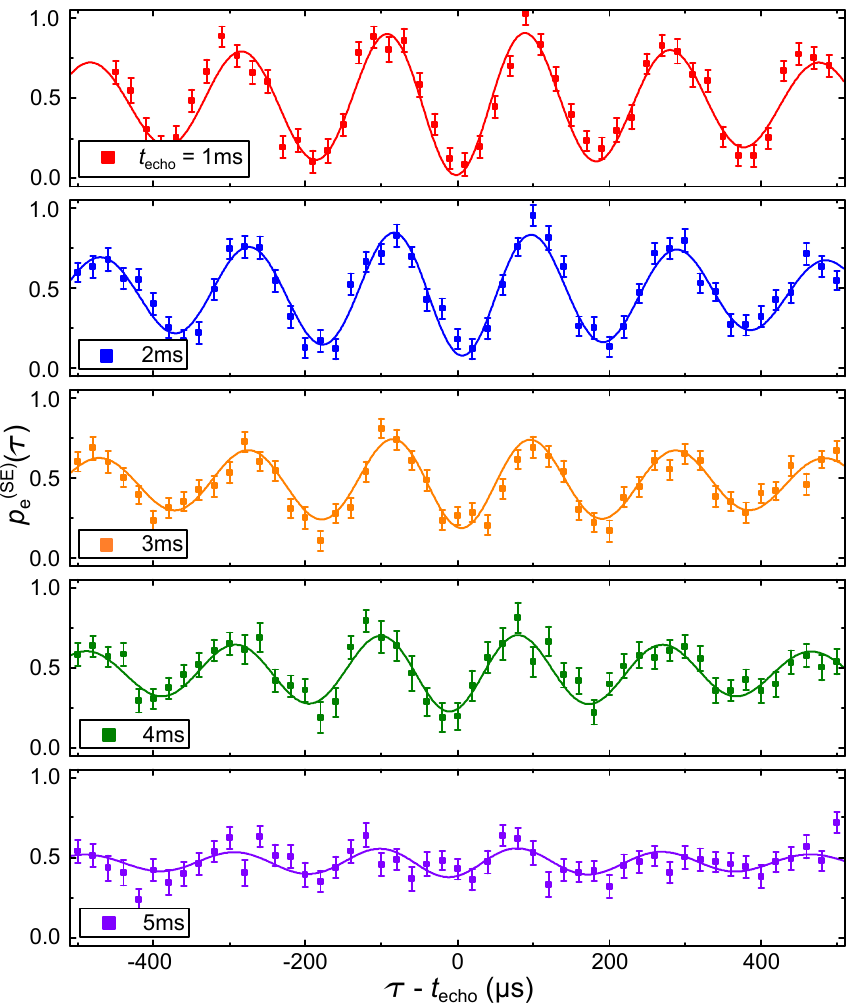}
	\caption{Probability $\pEc(\td)$ for an atom to be in $\ket{e}$ after a SE sequence. The data points correspond to an average over 80 experimental realizations (scaling factor $\alpha = 0.018$). The solid lines are the result of the global fitting routine. The legend in each panel indicates $\techo$.}
	\label{fig:echos} 
\end{figure}

In order to reverse the dephasing caused by the inhomogeneous broadening of the transition, the well-known spin echo (SE) technique is employed: A $\pi$-pulse is applied in between the two $\pi/2$-pulses at a time $\techo/2$ after the first $\pi/2$-pulse. In Fig.~\ref{fig:echos}, we plot the measured probability $\pEc(\td; \techo)$ of detecting an atom in $\ket{e}$. Spin echoes are apparent even for $\techo \gg \ttwos$.

The amplitude of the echo signal in Fig.~\ref{fig:echos} decreases within a characteristic time of a few milliseconds, indicating the presence of an irreversible dephasing mechanism. A general way to model the latter relies on the multiplication of the coherences (off-diagonal elements) of the density matrix $\rho$ by a time-dependent factor $0\leq C(t) \leq 1$ ($C(0) = 1$). This is done by introducing the superoperator $\decay(t)$ defined by
\begin{align}
\bra{i}\decay(t)\rho\ket{j} &= \left\{\begin{array}{l@{,\;}l} \rho_{ij} & i=j \\
C(t)\rho_{ij} & i\neq j 																												\end{array}\right. \mathrm{for}\, (i,j)\in\{e,g\}^2~.
\end{align}
The function $C(t)$ thus provides a measure of the coherence of the superposition~\cite{Cywinski2008, Sagi10}. The atomic spin density matrix at the end of the SE sequence reads
\begin{eqnarray}
	\se\rho(t;n) &=& \sesub{\boldsymbol{\seqOp}}(t;n) \inirho \nonumber\\	&&\hspace{-1.5cm}=\op_{\pi/2}\left[\decay(t)\left(\sesub{\evol}(t;n)\,\inirho\,\sesub{\evol}^\dag(t;n)\right)\right]\op_{\pi/2}^\dag,\label{eq:defRhoSE} \\
	\sesub{\evol}(t;n) &=&\free{t-\frac{\techo}{2};n}\op_\pi\free{\frac{\techo}{2};n}\op_{\pi/2}.\nonumber 
\end{eqnarray}
The evolution operator $\ramsub\seqOp$ in Eq.~\eqref{eq:defRho} has been replaced by the evolution superoperator $\sesub{\boldsymbol{\seqOp}}$. Moreover, $\decay(t)$ is assumed to commute with the operators $\evol_0$ and $\op_\pi$, which is fulfilled for ideal $\pi$-pulses and a good approximation in the case of small imperfections. For simplicity, we take $C(t)$ to be constant within the measurement time intervals  $|\td - \techo|\leq 0.5\,\milli\second$ and, thus, replace $\decay(t)$ in Eq.\eqref{eq:defRhoSE} by $\decay(\techo)$:
\begin{align}
	\se\rho(t;n) &= \label{eq:RhoCst}\\	&\hspace{-1.5cm}\op_{\pi/2}\left[\decay(\techo)\left(\sesub{\evol}(t;n)\,\inirho\,\sesub{\evol}^\dag(t;n)\right)\right]\op_{\pi/2}^\dag~. \nonumber
\end{align}
From this expression, the transfer probability $\pEc(\td;\techo)$ is calculated as in Eq.~\eqref{eq:idealProb}.

The solid lines in Fig.~\ref{fig:RabiRamsey}a, b, and in Fig.~\ref{fig:echos} are the result of a global fit of our model to the Ramsey fringes as well as to the SE signals. Rabi oscillations recorded over $\sim 5\,\milli\second$ are also included in the global fit. The fitting procedure disregards vibrational levels beyond $n_\mathrm{max} = 70$, which is well-justified for typical temperatures of the ensemble in the nanofiber-based trap. Imperfections of the $\pi$- and $\pi/2$-pulses due to the dependence of the rotation angle on $n$ via $\dls(n)$ are included. Furthermore, a phenomenological phase of the second $\pi/2$-pulse that accounts for a drift of the mean clock transition frequency during the sequence due to heating (see below) is added as a fit parameter. The other fit parameters are the initial temperature $T_0$, the MW detuning $\dmw$ in the Ramsey and SE experiments, the resonant Rabi frequency $\Omega_0$, the coefficients $C(\techo)$, and the ratio $\nato=N_{\rm tot}/\ntot$. The experimental data and the theory shown in Figs.~\ref{fig:RabiRamsey} and~\ref{fig:echos} agree very well. Moreover, the obtained initial temperature of $T_0=(71\pm4)\,\micro\kelvin$ is in reasonable agreement with what is expected from independent measurements of temperature and heating rate~\cite{Vetsch12,VetschPHD}. We note that the error bar of $T_0$ is only statistical and does not include possible systematic deviations due to, e.g., the finite accuracy of the calculation of the differential light shifts. The quality of the global fit together with the realistic range of the obtained fit parameters lead us to conclude that the measured reversible dephasing time $\ttwos$ is, indeed, a result of the joint effect of the finite initial temperature of the trapped atoms and the position-dependent differential light-shift of the clock transition. 

\begin{figure}
	\includegraphics{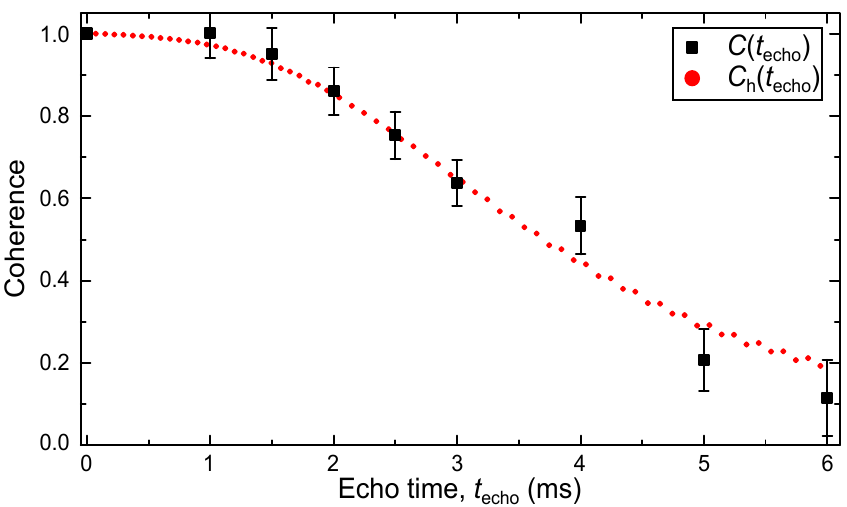}
	\caption{Irreversible decay of the atomic ground state coherence. Black squares: fitted parameters $C(\techo)$. The error bars are obtained by the standard error propagation method. Red dots: numerical simulation results obtained with a heating rate $\gamma\sim3\,\milli\kelvin\per\second$ and an initial temperature $T_0=71\,\micro\kelvin$.}
\label{fig:Coherence}
\end{figure}

We now model the decay of the spin echo amplitude, quantified by the fitted parameters $C(\techo)$, plotted in Fig.~\ref{fig:Coherence}. It has been shown previously that the heating rate of atoms in our nanofiber-based trap currently lies in the mK/s-range~\cite{VetschPHD}. We infer the effect of heating on the coherence of the trapped atoms and the resulting transfer probabilities $\pEc(\td;\techo)$ by calculating the time evolution of the density matrix $\wp$ of the joint system, consisting of the atomic spin and the vibrational phonons. We numerically solve the master equation $\ud \wp / \ud t = \mathcal{L}\wp$ written in the Lindblad form~\cite{Haroche2006}, with the initial state $\wp(0)$ corresponding to the tensor product of $\inirho=\braket{g}$ and the thermal state $\sum_n P(n,T_0)\braket{n}$. The Liouvillian $\mathcal{L}$ is given by
\begin{align}
\mathcal{L}\wp &= -i[H_0,\wp]  \label{eq:Liou}\\
&+ \kappa \sum_{\mu=+,-}\left[L_\mu\wp L_\mu^\dag-\frac{1}{2}\left(L_\mu^\dag L_\mu \wp + \wp L_\mu^\dag L_\mu\right)\right], \nonumber
\end{align}
where $H_0 = (\dls(\hat{n})-\dmw)\sigma_z/2$ is the free evolution Hamiltonian, the operator $L_- = {a}$ ($L_+ = {a}^\dag$) is the phonon annihilation (creation) operator, and $\hat{n} = {a}^\dag {a}$ is the phonon number operator. This Liouvillian describes the process of the atoms being heated due to a jitter of the center of the trap~\cite{Savard1997}. From the resulting $\pEc(\td;\techo)$, we deduce the expected time evolution of the coherences $\cheat(\techo)$. Here, in view of the typical temperatures of the atomic ensemble and in order to reduce computation time, the phonon Hilbert space is truncated at $n_\mathrm{max} = 30$. Figure~\ref{fig:Coherence} shows the results obtained for $\kappa = 350\,\second^{-1}$, corresponding to a heating rate of $\gamma\sim 3\,\milli\kelvin\per\second$ which is close to the one reported in~\cite{VetschPHD}, and $T_0=71\micro\kelvin$, taken from the global fit of the data in Figs.~\ref{fig:RabiRamsey} and~\ref{fig:echos}. The theoretical prediction matches very well with values of $C(\techo)$ extracted from the experimental data. We therefore conclude that for our current experimental parameters, the irreversible dephasing is dominated by heating of the atoms. We define the irreversible dephasing time $\ttwo$ as the half width at half maximum of $\cheat(\techo)$ and obtain $\ttwo=3.7\,\milli\second$.

Summarizing, we showed that ground state coherence times in the range of milliseconds can be achieved in our nanofiber-based atom trap~\cite{Vetsch10} with atoms at a distance of only $\sim 200$~nm from the hot silica fiber surface~\cite{arxiv_Wuttke12}. Currently, the coherence times are limited by the finite initial temperature of the atomic ensemble and the heating rate. Both can in principle be further decreased by technical measures like Raman side-band cooling~\cite{Kaufman12,Thompson13}. This constitutes a decisive result towards establishing nanofiber-based quantum interfaces as practical building blocks in an optical fiber quantum network. In particular, they pave the way towards the realization of fully fiber-integrated quantum memories which could also be operated as highly efficient photon number-resolving detectors~\cite{Peyronel12b}. 

We thank Fam Le Kien for helpful discussions. This work was supported by the Volkswagen Foundation and the European Science Foundation. We acknowledge financial support by the FWF project CoQuS No.~W1210-N16.


\bibliography{Ramsey2}

\end{document}